\documentclass[a4paper,fleqn,breaklinks]{cas-dc}

\usepackage[numbers]{natbib}
\usepackage{url}
\usepackage{caption}
\usepackage{graphicx}

\newcommand{\blue}[1]{\textcolor{blue}{#1}}

\def\tsc#1{\csdef{#1}{\textsc{\lowercase{#1}}\xspace}}
\tsc{WGM}
\tsc{QE}
\tsc{EP}
\tsc{PMS}
\tsc{BEC}
\tsc{DE}

\begin{document}
	\let\WriteBookmarks\relax
	\def\floatpagepagefraction{1}
	\def\textpagefraction{.001}
	\shorttitle{COVID-19 Misinformation on Twitter}
	\shortauthors{Shahi et~al.}
	
	\title[mode = title]{An Exploratory Study of COVID-19 Misinformation on Twitter}
	\tnotemark[1]
	
	\tnotetext[1]{The work presented in this document results from the Horizon 2020 Marie Skłodowska-Curie project \emph{RISE\_SMA} funded by the European Commission.}
	
	\author[1]{Gautam Kishore Shahi}[orcid=0000-0001-6168-0132]
	\ead{gautamshahi16@gmail.com}
	\address[1]{University of Duisburg-Essen, Germany}
	\credit{Data Collection, Data curation, Investigation, Methodology, Software, Visualization, Writing -- original draft, Writing -- review \& editing}
	
	\author[2]{ Anne Dirkson}[orcid = 0000-0002-4332-0296]
	\address[2]{LIACS, Leiden University, Netherlands}
	\ead{a.r.dirkson@liacs.leidenuniv.nl}
	\credit{Data curation, Investigation, Methodology, Software, Visualization, Writing -- original draft, Writing -- review \& editing}
	
	\author[3]{ Tim A. Majchrzak}[orcid = 0000-0003-2581-9285]
	\cormark[2]
	\ead{timam@uia.no}
	\address[3]{University of Agder, Norway}
	\credit{Conceptualization, Funding acquisition, Methodology, Project administration, Writing -- original draft, Writing -- review \& editing}
	
	\cortext[2]{Corresponding author}
	
	\begin{abstract}
		During the COVID-19 pandemic, social media has become a home ground for misinformation. To tackle this infodemic, scientific oversight, as well as a better understanding by practitioners in crisis management, is needed. We have conducted an exploratory study into the propagation, authors and content of misinformation on Twitter around the topic of COVID-19 in order to gain early insights. We have collected all tweets mentioned in the verdicts of fact-checked claims related to COVID-19 by over 92 professional fact-checking organisations between January and mid-July 2020 and share this corpus with the community. This resulted in 1\,500 tweets relating to 1\,274 false and 276 partially false claims, respectively. Exploratory analysis of author accounts revealed that the verified twitter handle(including Organisation/celebrity) are also involved in either creating (new tweets) or spreading (retweet) the misinformation. Additionally, we found that false claims propagate faster than partially false claims. Compare to a background corpus of COVID-19 tweets, tweets with misinformation are more often concerned with discrediting other information on social media. Authors use less tentative language and appear to be more driven by concerns of potential harm to others. Our results enable us to suggest gaps in the current scientific coverage of the topic as well as propose actions for authorities and social media users to counter misinformation.
	\end{abstract}
	
	\begin{highlights}
		\item A very timely study on misinformation on the COVID-19 pandemic
		\item Synthesis of social media analytics methods suitable for analysis of the infodemic
		\item Explaining what makes COVID-19 misinformation distinct from other tweets on COVID-19
		\item Answering where COVID-19 misinformation originates from, and how it spreads
		\item Developing crisis recommendations for social media listeners and crisis managers
	\end{highlights}
	
	
	\maketitle
	
	\section{Introduction}
	The COVID-19 pandemic is currently spreading across the world at an alarming rate~\cite{world2020coronavirus}. It is considered by many to be the defining global health crisis of our time~\cite{sohrabi2020world}. 
	As WHO Director-General Tedros Adhanom Ghebreyesus proclaimed at the Munich Security Conference on 15 February 2020, "We're not just fighting an epidemic; we're fighting an \emph{infodemic} ".~\cite{Zarocostas2020}. It has even been claimed that the spread of COVID-19 is \emph{supported} by misinformation~\cite{Garrett2020}. The actions of individual citizens guided by the quality of the information they have at hand are crucial to the success of the global response to this health crisis. By 18 July 2020, the International Fact-Checking Network (IFCN)~\cite{Poynter2020} uniting over 92 fact-checking organisations unearthed over 7\,623 unique fact checked articles regarding the pandemic.
	However, misinformation does not only contribute to the spread: misinformation might bolster fear, drive societal disaccord, and could even lead to direct damage -- for example through ineffective (or even directly harmful) medical advice or through over- (e.g. hoarding) or underreaction (e.g. deliberately engaging in risky behaviour)~\cite{Pennycook2020}.
	
	Misinformation on COVID-19 appears to be spreading rapidly on social media~\cite{Zarocostas2020}. Similar trends were seen during other epidemics, such as the recent Ebola ~\cite{Oyeyemi2014}, yellow fever~\cite{Ortiz-Martinez2017} and Zika~\cite{miller2017} outbreaks. This is a worrying development as even a single exposure to a piece of misinformation increases its perceived accuracy~\cite{pennycook2018prior}. In response to this infodemic, the WHO has set up their own platform \emph{MythBusters} that refutes misinformation~\cite{WHO2020} and is urging tech companies to battle fake news on their platforms~\cite{BBC}.\footnote{At the same time, the WHO itself faces criticism regarding how it handles the crisis, among others regarding the dissemination of information from member countries~\cite{GuardianWHO}.} Fact-checking organisations have united under the IFCN to counter misinformation collaboratively, as individual fact-checkers like Snopes are being overwhelmed~\cite{BusinessInsider2020}. 
	
	There are many pressing questions in this uphill battle. So far\footnote{The number of articles taking COVID-19 as an example, case study, or even directly as the main theme is steadily growing. In particular, there is a high number of preprints; how many of those will eventually make it to the scientific body of knowledge remains to be seen. Any mentioning of closely related work in such articles -- including ours -- must be seen as a snapshot, as moments after submission likely more work is uploaded elsewhere.}, five studies have investigated the magnitude or spread of misinformation on Twitter regarding the COVID-19 pandemic \cite{Cinelli, Kouzy2020, Gallotti2020, Singh2020, Yang2020}. However, two of these studies either investigated a very small subset of claims \cite{Singh2020} or manually annotated a small subset of Twitter data \cite{Kouzy2020}. The remaining studies used the reliability of the cited sources to identify misinformation automatically \cite{Gallotti2020, Cinelli, Yang2020}. Although such source-based approaches are popular and allow for a large-scale analysis of Twitter data \cite{Allen2020, Zollo2017, Bovet2019, Grinberg2019, Shao2018}, the reliability of news sources remains a subject of considerable disagreement~\cite{Vosoughi2018, Pierri2020}. Moreover, source-based classification misclassifies misinformation propagated by generally reliable mainstream news sources \cite{Allen2020} and misses misinformation generated by individuals, e.g. by Donald Trump or unofficial sources such as recently emerging web sites. According to a recent report for the European Council by Wardle and Derakhshan \cite{Wardle2017}, the latter is increasingly prevalent.

In our study, we employ an alternative, complementary approach also used by \cite{Vosoughi2018, Monti2019, Rosenfeld2020}; We rely on the verdicts of professional fact-checking organisations which manually check each claim. This does limit the scope of our analysis due to the bottleneck of verifying claims but avoids the limitations of a source-based approach, thereby complementing previous work. Furthermore, none of the previous studies has investigated how the language use of COVID-19 misinformation differs from other COVID-19 tweets or which Twitter accounts are associated with the spreading of COVID-19 misinformation. Although there have already been some indications that bots might be involved \cite{Gallotti2020, Ferrara2020, Yang2020}, the majority of posts is generated by accounts that are likely to be human \cite{Yang2020}. 

We thus conduct an exploratory analysis into (1) the Twitter accounts behind COVID-19 misinformation, (2) the propagation of COVID-19 misinformation on Twitter, and (3) the content of incorrect claims on COVID-19 that circulate on Twitter. We decided to work exploratory because too little is known about the topic at hand to tailor either a purely quantitative or a purely qualitative study.

The exploration of the phenomena with the aim of rapid dissemination of results combined with the demand for academic rigour make our article somewhat uncommon in nature. We, therefore, explicate our three contributions.
First, we present a synthesis of social media analytics techniques suitable for the analysis of the COVID-19 infodemic. We believe this to be a starting point for a more structured, goal-oriented approach to mitigate the crisis on the go -- and to learn how to decrease negative effects from misinformation in future crisis as they unfold. Second, we contribute to the scientific theory with first insights into how COVID-19 misinformation differs from other COVID-19 related tweets, which it originates from, and how it spreads. This should pose the foundation for drawing a research agenda. Third, we provide the first set of recommendations for practice. They ought to directly help social media managers of authorities, crisis managers, and social media listeners in their work.

In Section~\ref{sec:rw}, we provide the academic context of our work in the field of misinformation detection and propagation. In Sections~\ref{sec:meth} and~\ref{sec:data}, we elaborate on our data collection process and methodology, respectively. We then present the experimental result in Section~\ref{sec:exp}, followed by discussing these results and providing recommendations for organisations targeting misinformation in Section~\ref{sec:dis}. Finally, we draw a conclusion in Section~\ref{sec:conclusion}.

\section{Background \label{sec:rw}}
In this section, we describe the background of misinformation, propagation of misinformation, rumours detection, and the impact of fact-checking.

\subsection{Defining misinformation}
Within the field there is no consensus on the definition for misinformation \cite{Pierri2020}. We define misinformation broadly as circulating information that is \emph{false} \cite{Zubiaga2017a}. The term misinformation is more commonly used to refer specifically to when false information is shared \emph{accidentally}, whereas disinformation is used to refer to false information shared \emph{deliberately}~\cite{hernon1995}. In this study, we do not make claims about the intent of the purveyors of information, whether accidental or malicious. Therefore, we pragmatically group \textit{false} information regardless of intent. In line with recommendations by Wardle and Derakhshan \cite{Wardle2017}, we avoid the polarised and inaccurate term \emph{fake news}. 

Additionally, there is no consensus on when a piece of information can be considered false. According to seminal work by del Vicario \textit{et al.} \cite{Vicario2016} it is ``the possibility of verification rather than the quality of information'' that is paramount. Thus, verifiability should be considered key to determining falsity. To complicate matters further, claims are not always wholly false or true, but there is a scale of accuracy \cite{Wardle2017}. For instance, claims can be mostly false with elements of truth. Two examples in this category are images that are miscaptioned and claims to omit necessary background information. In our work, we name such claims \emph{partially false}.

We rely on the manual evaluations of fact-checking organisations to determine which information is (partially) false (see Section \ref{sec:classdef} for details on the conversion of manual evaluations from fact-checkers). We make the distinction between false and partially false for two reasons. First, other researchers have proposed a scale over a hard boundary between false and not false, as illustrated above. Second, it needs to be assessed, whether completely and partially false information is perceived differently. We expect the \emph{believability} of partially false information to be higher. It may be more challenging for users to recognise claims as false when they contain elements of truth, as this has found to be the case even for professional fact-checkers~\cite{Lim2018}.

The comparison between partially and completely false claims thus enables us to attain better insight into differences in their spread. It is crucial for fact-checking organisations and governments battling misinformation to understand better how to sustain \emph{information sovereignty}~\cite{HICSS2018SMR}. In an ideal setting, people would always check facts and employ scientific methods. In a realistic setting, they would at least be mainly drawn to information coming from fact-based sources which work ethically and without a hidden agenda. Authorities such as cities (local governments) ought to be such sources~\cite{SCC2019}. 

\subsection{Identifying rumours on Twitter}
Rumours are ``circulating pieces of information whose veracity is yet to be determined at the time of posting''~\cite{Zubiaga2017a}. Misinformation is essentially a false rumour that has been debunked. Research on rumours is consequently closely related, and the terms are often used interchangeably. 

Rumours on social media can be identified through top-down or bottom-up sampling~\cite{Zubiaga2017a}. A top-down strategy use rumours which have already been identified and fact-checked to find social media posts related to these rumours. This has the disadvantage that rumours that have not been included in the database are missed. Bottom-sampling strategies have emerged more recently and are aimed at collecting a wider range of rumours often prior to fact-checking. This method was first employed by by~\cite{Zubiaga2016}. However, manual annotation is necessary when using a bottom-up strategy. Often journalists with expertise in verification are enlisted since crowd-sourcing will lead to credibility perceptions rather than ground truth values. The exhaustive verification may be beyond their expertise~\cite{Zubiaga2017a}.

In this study, we employ a top-down sampling strategy relying on the work of on Snopes.com and over 91 different fact-checking organisations organised under the CoronaVirusFacts/ DatosCoronaVirus alliance run by the Poynter Institute. We included all misinformation (see Section \ref{sec:classdef}) around the topic of COVID-19, which include a Tweet ID. A similar approach was used by Jiang \textit{et al.}~\cite{Jiang2018} with Snopes.com and Politifact and by~\cite{Vosoughi2018} using six independent fact-checking organisations. 

\subsection{Misinformation propagation}
To what extent information goes \emph{viral} is often modelled using epidemiological models originally designed for biological viruses~\cite{Goel2016, Serrano}. The information is represented as an `infectious agent' that is spread from `infectives' to `susceptibles' with some probability. This method was also employed by~\cite{Cinelli} for the propagation of information to study how infectious information on COVID-19 is on Twitter. They found that the basic reproductive number $R_0$, i.e. the number of infections due to one infected individual for a given period, is between 4.0 to 5.1 on Twitter, indicating a high level of `virality' of COVID-19 information in general.\footnote{This, curiously, means that misinformation on the new coronavirus has higher infectivity than the virus itself \cite{liu2020reproductive,zhang2020estimation}.} Additionally, they found the overall magnitude of COVID-19 misinformation on Twitter to be around 11\%. They also investigated the relative amplification of reliable and unreliable information on Twitter and found it to be roughly equal. Similarly, Yang \textit{et al.}~\cite{Yang2020} found that the volume of tweets linking to low-credibility information was compared to the volume of links to the New York Times and Centre for Disease Control and Prevention (CDC).

Other researchers have modelled information propagation on Twitter using the retweet (RT) trees, i.e. asking \emph{who retweets whom}? Various network metrics can then be applied to quantify the spread of information such as the depth (number of retweets by unique users over time), size (number of total users involved) or breadth (number of users involved as a certain depth)~\cite{Vosoughi2018}. These measures can also be considered over time to understand how propagation fluctuates. Additionally, these networks can be used to investigate the role of bots in the spreading of information \cite{Shao2018}. Recent studies \cite{Monti2019, Pierri2020} have also shown the promise of propagation-based approaches for precise discrimination of fake news on social media. In fact, it appears that aspects of tweet content can be predicted from the collective diffusion pattern \cite{Rosenfeld2020}.

An advantage of this approach compared to epidemiological modelling, is that it does not rely on the implicit assumption that propagation is driven largely if not exclusively by peer-to-peer spreading~\cite{Goel2016}. However, viral spreading is not the only mechanism by which information can spread: Information can also be spread by broadcasting, i.e. a large number of individuals receive information directly from one source. Goel et al.~\cite{Goel2016} introduced the measure of \emph{structural virality} to quantify to what extent propagation relies on both mechanisms. 

\subsection{The impact of fact-checking}
Previous research on the efficacy of fact-checking reveals the corrections often do not have the desired effect and misinformation resists debunking \cite{Zollo2017}. Although the likelihood of sharing does appear to drop after a fact-checker adds a comment revealing this information to be false, this effect does not seem to persist on the long run~\cite{Friggeri2014}. In fact, 51.9\% of the re-shares of false rumours occur after this debunking comment. This may, in part, be due to readers not reading all the comments before re-sharing. Complete retractions of the misinformation are also generally ineffective, despite people believing, understanding and remembering the retraction~\cite{Lewandowsky2012}. \emph{Social reactance}~\cite{brehm} may also play a role here: people do not like being told what to think and may reject authoritative retractions. Three factors that do increase their effectiveness are (a) repetition, (b) warnings at the initial exposure and (c) corrections that tell an alternate story that does not leave behind an unexplained gap~\cite{Lewandowsky2012}. 

Twitter users also engage in debunking rumours. Overall, research supports the idea that the Twitter community debunks inaccurate information through self-correction~\cite{Zubiaga2017a, Mendoza2010}. However, self-correction can be slow to take effect~\cite{Procter2013} and interaction with debunking posts can even lead to an increasing interest in conspiracy-like content \cite{Zollo2017}. Moreover, it appears that in the earlier stages of a rumour circulating Twitter users have problems differentiating between true and false rumours~\cite{Zubiaga2016}. This includes users of the high reputation such as news organisations who may issue corrective statements at a later date if necessary. This underscores the necessity of dealing with newly emerging rumours around crises like the outbreak of COVID-19. 

Yet, these corrections also do not always have the desired effect. Fact-checking corrections are most likely to be tweeted by strangers but are more likely to draw user attention and responses when they come from friends~\cite{Hannak2014}. Although such corrections do elicit more responses from users containing words referring to facts, deceit (e.g. fake) and doubt, there is an increase in the number of swear words~\cite{Jiang2018}, too. Thus, on the one hand, users appear to understand and possibly believe the rumour is false. On the other hand, swearing likely indicates backfire~\cite{Jiang2018}: an increase in negative emotion is symptomatic of individuals clinging to their own worldview and false beliefs. Thus, corrections have mixed effects that may depend in part on who is issuing the correction. 

\section{Data collection \& preprocessing\label{sec:data}}
In this section, we describe the steps involved in the data collection and filtering the tweets for analysis. We have used two datasets for our study. The first dataset are the tweets which have been mentioned by fact-checking websites and are classified as false or partially false and the second dataset consists of COVID-19 tweets collected from publicly available corpus TweetsCOV19 \footnote{\url{https://data.gesis.org/tweetscov19/} (January-April 2020)} and in-house crawling from May-July 2020. A detailed description of data collection process explain in section \ref{sec:datacol}.

\subsection{Data collection \label{sec:datacol}}
For our study, we gathered the data from two different sources. The first data set consists of false or partially false tweets from the fact-checking websites. The second is a random sample of tweets related to COVID-19 from the same period.

\subsubsection{Dataset I -- Misinformation Tweets}
We used an automated approach to retrieve tweets with misinformation. First, we collected the list of fact-checked news articles related to the COVID-19 from Snopes~\cite{snopes} and Poynter~\cite{Poynter2019} from 04-01-2020 to 18-07-2020. We collected 7\,623 fact-checked articles using the approach mentioned in \cite{shahi2020fakecovid}. We used Beautifulsoup~\cite{richardson2007beautiful} to crawl the content of the news articles and prepared a list of news articles which collected the information like title, the content of the news article, name of the fact-checking website, location, category (e.g. False, Partially False) of fact-checked claims.

\begin{figure*}[pos=t]
	\centering
	\includegraphics[width=0.96\textwidth]{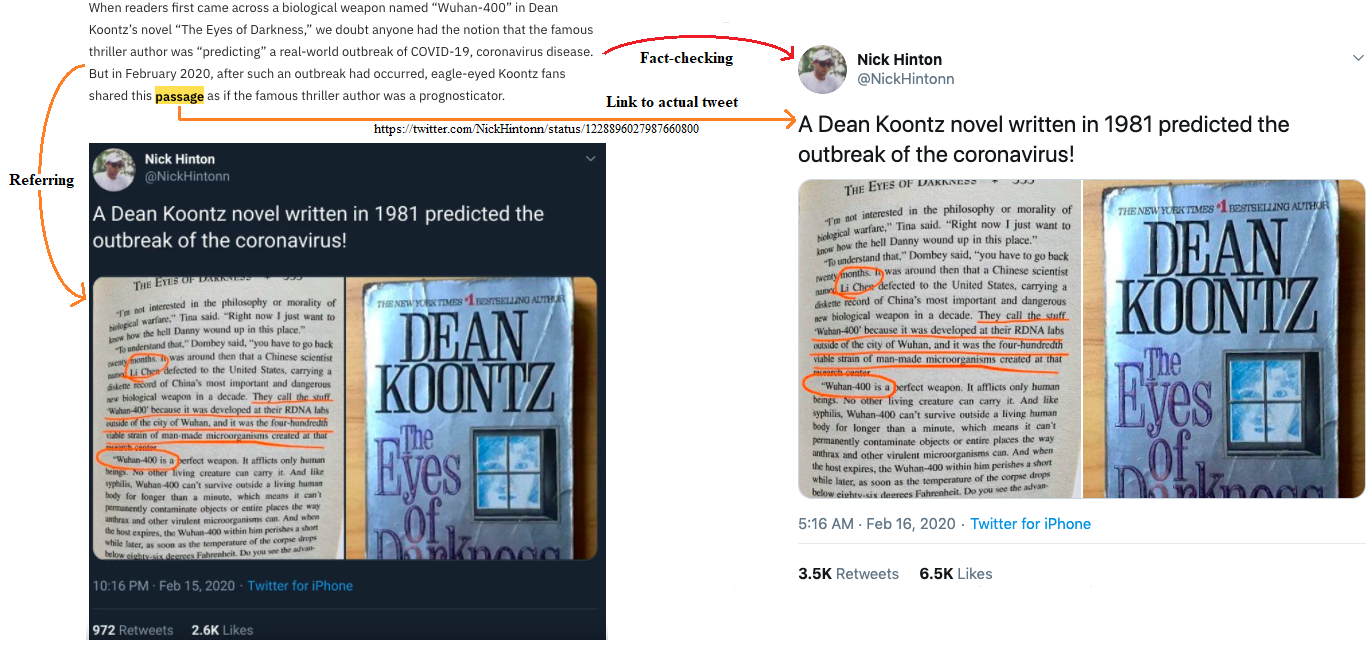}
	\caption{Illustration of data collection method- Extraction of social media link (Tweet Link) on the fact checked article and fetching the relevant tweets from Twitter (screenshots from \cite{examplesTweetHinton,examplesDebunkEvon}). }
	\label{fig:dc}
\end{figure*}

To find the misleading posts on COVID-19 on Twitter, we crawled the content of the news article using Beautifulsoup and looked for the article, which is referring to Twitter. In the HTML Document Object Model(DOM), we looked for all anchor tags <a> which defines a hyperlink. We filter the anchor tag which contains keyword `twitter' and `status' because each tweet message is linked with the Uniform Resource Locator(URL) in the form of \url{https://twitter.com/statuses/ID}. From the collected URLs, we fetched the ID, where the ID is the unique identifier for each tweet. An illustration of the overall workflow for fetching tweets mentioned in the fact-checked articles is shown in Figure~\ref{fig:dc}.
We collected a total of 3\,053 Tweet IDs from 7\,623 news articles. The timestamps of these tweets are between 14-01-2020 and 10-07-2020. After removing the duplicates tweets, we got 1\,565 tweets for our analysis which is further filtered based on its category as discussed in section \ref{sec:classdef}. We further categorise the tweet ID into four different classes as mentioned in section \ref{sec:classdef}.

From the Tweet ID generated in the above step, we used tweepy \cite{tweepy}, a python library for accessing the Twitter API. Using the library, we fetched the tweet and its description such as created\_at, like, screen name, description, and followers.

\subsubsection{Dataset II -- Background corpus of COVID-19 Tweets \label{sec:back}}
To understand how the misinformation around COVID-19 is distinct from the other tweets on this topic, we created a background corpus of 163\,096 English tweets spanning the same time period (14 January until 10 July) as our corpus of misinformation. We randomly selected 1\,000 tweets per day and all tweets if fewer than 1\,000 tweets were available. For January until April, we used the publicly available corpus TweetsCOV19\footnote{\url{https://data.gesis.org/tweetscov19/}. We attempted to retrieve tweet content using the Twitter API. As some tweets were no longer available, this resulted in 92\,095 tweets. TweetsCOV19 spans until April 2020  so, for May to July, we used our own keyword-based crawler using Twitter4J, resulting in a total of 71\,000 tweets for this time span. Specifically, we used several hashtags, including \#Coronavirus, \#nCoV2019, \#WuhanCoronovirus, \#WuhanVirus, \#Wuhan, \#CoronavirusOutbreak, \\
	\#Ncov2020, \#coronaviruschina, \#Covid19, \#covid\-19, \\ \#covid\_19, \#sarscov2, \#covid, \#cov, and \#corona. We merged both data sets for our study.}

\subsubsection{Retweets and Account details}
In this section, we describe the methods used for the crawling of retweets and the retrieval of details of author accounts. 
\paragraph{\textbf{Retweets}} Usually, Fake news spreads on social media immediately after sharing the post on social media. We wanted to analyse the difference in the propagation of false and partially false tweets. We fetched all the retweet using the python library Twarc~\cite{twarc}. Twarc is a command-line tool for collecting Twitter data in JSON\footnote{https://www.json.org/} format. 

Our main goal to detect the difference in the propagation speed for false and partially false tweets, so we crawled the retweets for the data set I only.

\paragraph{\textbf{User account details}} 
From the Twitter API, we also gathered the account information: favourites count (number of likes gained), friends count (number of accounts followed by the user), follower count (number of followers this account currently has), account age (number of days from account creation date to 31-12-2020, the time when discussion about COVID-19 started around the world), a profile description, and user location. We used this information for both classifying the popular accounts and for bot detection. 

\subsection{Defining classes for misinformation \label{sec:classdef}} 
Discounting differences in capitalisation, our data originally contained 18 different verdict classes provided by the 92 fact-checking websites, i.e. Snopes and 91 organisations in the International Fact Checking Network (IFCN). In Table~\ref{tab:NewsCategorization}, we provide an overview of the verdict categories that were included or excluded in our study along with our categorisation and the original, more granular categorisation by fact-checkers. Since each fact-checking organisation has its own set of verdicts and Poynter has not normalised these, manual normalisation is necessary. Following the practice of~\cite{Vosoughi2018}, we normalised verdicts by manually mapping them to a score of 1 to 4 (1=`False', 2=`Partially False', 3=`True', 4=`Others') based on the definitions provided by the fact-checking organisations. Our definition for the four categories are as follows- \newline
\textbf{False}: Claims of an article are untrue. \newline
\textbf{Partially False}: Claims of an article are a mixture of true and false information. The article contains partially true and partially false information, but it can not be considered as 100\% true. It includes articles of type, partially false, partially true, mostly true, miscaptioned, misleading etc. \newline
\textbf{True}: This rating indicates that the primary elements of a claim are demonstrably true. \newline
\textbf{Other}: An article that cannot be categorised as true, false or partially false due to lack of evidence about its claims. This category includes articles in dispute and unproven articles.

As we are specifically interested in misinformation, we considered only the false and partially false category. We also excluded claims with verdicts that did not conform to this scale, e.g. sarcasm, unproven claims and disputed claims. From 1\,565 tweets collected, 1\,500 are used for our study -- 1\,274 false and 226 partially false claims. The data used in our work is available through GitHub\footnote{\url{https://github.com/Gautamshahi/Misinormation_COVID-19}}.

\begin{table*}[pos=p!]
	\caption{Normalisation of original categorisation by the fact checking web sites}\label{tab:NewsCategorization}
	\begin{tabular}{ p{0.85cm} p{1.9cm} p{2.15cm} p{10,58cm}  } 
		
		\toprule
		Included (y/n) & Our rating     & Fact-checker rating & Definition given by fact-checker      \\
		
		\bottomrule
		y              & False           & False               & The checkable claims are all false.             \\                                                                                                              
		y              & Partially false & Miscaptioned        & This rating is used with photographs and videos that are “real” (i.e., not the product, partially or wholly, of digital manipulation) but are nonetheless misleading because they are accompanied by explanatory material that falsely describes their origin, context, and/or meaning.                                                                                                                                                         \\
		y              & Partially false & Misleading          & Offers an incorrect impression on some aspect(s) of the science, leaves the reader with false understanding of how things work, for instance by omitting necessary background context.                                                                                                                                                                                                                                                         \\
		n              & Others               & Unsupported/ Unproven            & This rating indicates that insufficient evidence exists to establish the given claim as true, but the claim cannot be definitively proved false. This rating typically involves claims for which there is little or no affirmative evidence, but for which declaring them to be false would require the difficult (if not impossible) task of our being able to prove a negative or accurately discern someone else’s thoughts and motivations. \\
		y              & Partially false & Partially false     & {[}Translated{]} Some claims appear to be correct, but some claims can not be supported by evidence.                                                                                                                                                                                                                                                                                                                                            \\
		y              & False           & Pants on fire/ Two pinocchios        & The statement is not accurate and makes a ridiculous claim.                                                                                                                                                                                                                                                                                                                                                                                     \\
		y              & Partially false & Mostly false        & Mostly false with one minor element of truth.                                                                                                                                                                                                                                                                                                                                                                                                   \\
		n              & True             & True                & This rating indicates that the primary elements of a claim are demonstrably true.                                                                                                                                                                                                                                                                                                                                                               \\
		n              & Others               & Labeled Satire      & This rating indicates that a claim is derived from content described by its creator and/or the wider audience as satire. Not all content described by its creator or audience as ‘satire’ necessarily constitutes satire, and this rating does not make a distinction between 'real' satire and content that may not be effectively recognized or understood as satire despite being labelled as such.                                           \\
		n              & Others               & Explanatory         & "Explanatory" is not a rating for a checked article, but an explanation of a fact on its own                                                                                                                                                                                                                                                                                                                                                    \\
		y              & Partially false               & Mixture             & This rating indicates that a claim has significant elements of both truth and falsity to it such that it could not fairly be described by any other rating.                                                                                                                                                                                                                                                                                     \\
		y             & Partially false               & Mostly true         & Mostly accurate, but there is a minor error or problem.                                                                                                                                                                                                                                                                                                                                                                                         \\
		y              & Partially false & Misinformation/ Misattributed       & This rating indicates that quoted material (speech or text) has been incorrectly attributed to a person who didn't speak or write it.                                                                                                                                                                                                                                                                                                           \\
		n              & Others               & In dispute          & One can see the duelling narratives here, neither entirely incorrect. For that reason, we will leave this unrated.                                                                                                                                                                                                                                                                                                                               \\
		
		y              & False           & Fake                & {[}Rewritten generalized{]} Claims of an article are untrue                                                                                                                                                                                                                                                                                                                                                                                     \\
		n              & Others               & No rating   &  Outlet decided to not apply any rating after doing a fact checking.      \\
		
		y              & Partially false              & Partially True   &  Leaves out important information or is made out of context.      \\
		
		y              & Partially false              & Manipulations  &  [Translated] Article only showed part of an interview answer, and interview question has been phrased in a way that makes it easy to manipulate the answer.      \\

		\bottomrule
	\end{tabular}
	
\end{table*}
\subsection{Examples of misinformation}

\begin{figure}[pos=t!]
	\centering
	\includegraphics[width=0.95\linewidth]{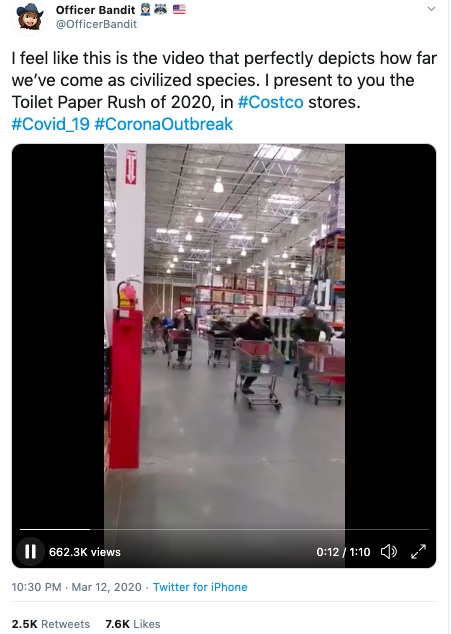}
	\caption{An example of misinformation of false category}
	\label{fig:ex1}
\end{figure}

\begin{figure}[pos=t!]
	\centering
	\includegraphics[width=0.95\linewidth]{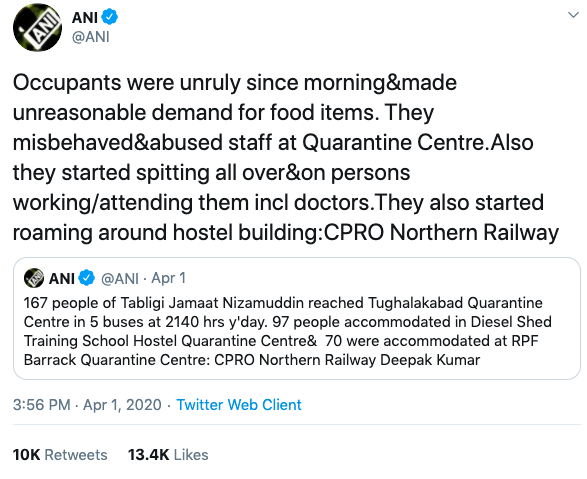}
	\caption{An example of Misinformation of partially false category}
	\label{fig:ex2}
\end{figure}

Figure~\ref{fig:ex1} and Figure~\ref{fig:ex2} display two randomly chosen examples of misinformation in the false and partially false category, respectively. The first is an example of a false claim, namely a tweet about the rumour that Costco had issued a recall of their toilet paper because they feared that it might contain COVID-19~\cite{false1}. The author states that people were running to the store to buy and then return the toilet paper after hearing the news. Later the claim was fact-checked by Snopes and found to be false, and no such recall had been announced by Costco~\cite{snopes1}. There were several other tweets making similar false claims.

An example of a tweet~\cite{pfalse1} containing partially false information was posted by the news company, ANI. It claimed that people quarantined from Tablighi Jamaat~\cite{ali2003islamic} misbehaved towards health workers and police staff. They were not following the rules of the quarantine centre and misbehaved the police. AFP~\cite{AFP2020b} found the claim to be misleading. The misleading claim is one subset of the claim that is normalised to partially false. The incident used in the claim was used from a past event in Mumbai during February 2020. Different Twitter handles circulated this misinformation. The claim was retweeted and liked by several users on Twitter. 

The first tweet about toilet paper is false because the tweet with old video was circulated with false information about the recall of the toilet paper. The second tweet about Tablighi Jamaat is considered partially false as the incident portraying a true incident that people of Tablighi Jamaat are not following the protocol, but it was re-purposed with a different video to support a false claim. So, if a claim is based on a false incident or information its called a false claim while if it's based on the true incident with a false message then its partially false.

\subsection{Preprocessing of tweets \label{sec:datasec}}
\begin{figure}
	\centering
	\includegraphics[width=.48\textwidth]{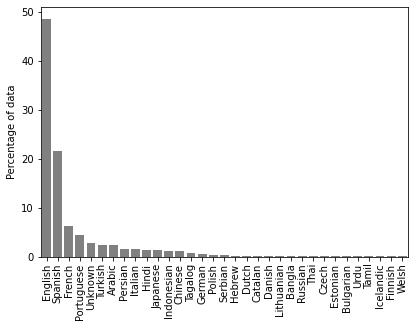}
	\caption{The language distribution of tweets with misinformation prior to translation of tweets}
	\label{fig:langdistHist}
\end{figure}

Originally, the data contained 32 known languages (according to Twitter -- see Figure \ref{fig:langdistHist}). We use the Google Translate API\footnote{https://cloud.google.com/translate/docs} to automatically detect the correct language and translate to English. Hereafter, tweets were lowercased and tokenised using NLTK\cite{loper2002nltk}. Emojis were identified using the emoji package \cite{emojiPython} and were removed for subsequent analyses. Mentions and URLs were also removed using regular expressions. Hashtags were not removed, as they are often used by twitter users to convey essential information. Additionally, sometimes they are used to replace regular words in the sentence (e.g. `I was tested for \#corona') and thus omitting them would remove essential words from the sentence. Therefore, we only remove the \# symbol from the hashtags.

\section{Method \label{sec:meth}}
In this section, we present our method for analysis and illustration of the extracted data. We follow a two-way approach.
In the first, we analyse the details of the user accounts involved in the spread of misinformation and propagation of misinformation (false or partially false data). In the second, we analyse the content. With both we investigate the propagation of misinformation on social media.

\subsection{Account categorisation}
\label{account}
In order to gain a better understanding of who is spreading misinformation on Twitter, we investigated the Twitter accounts behind the tweets. First, we analyse the role of bots in spreading misinformation by using a bot detection API to automatically classify the accounts of authors. Similarly, we analyse whether accounts are brands using an available classifier. Third, we investigate some some characteristics of the accounts that reflect their popularity (e.g. follower count).

\subsubsection{Bot detection}
A Twitter bot is a type of bot program which operate a Twitter account via the Twitter API. The pre-programmed bot autonomously performs some work such as tweeting, unfollowing, re-tweeting, liking, following or direct messaging other accounts. Shao et al.~\cite{shao2017spread} discussed the role of social bots in spreading the misininformation. Previous studies show there are several types of bots involved in social media such as "newsbots", "spambots", "malicious bot". Sometimes, newsbots or malicious bots are trained to spread the misinformation.   Caldarelli et al.~\cite{caldarelli2019role} discuss the role of bots in Twitter propaganda. To analyse the role of bots, we examined each account by using a bot detection API~\cite{davis2016botornot}.

\subsubsection{Type of account (brand or non-brand)}
Social media, such as microblogging websites, used for sharing information and gathering opinion on the trending topic. Social media has different types of user, organisation, celebrity or an ordinary user. We consider organisation, celebrity as a brand which has a big number of followers and catches more attention public attention. The brand uses a more professional way of communication, gets more user attention~\cite{sook2014brand} and have high reachability due to bigger follower network and retweet count. With a large network, a piece of false or partially false information spread faster compared to a normal account. We classify the account as a brand or normal users using a modified of TwiRole~\cite{li2018hybrid} a python library. We use profile name, picture, latest tweet and account description to classify the account.

\subsubsection{Popularity of account} 
Popular accounts get more attention from users, so we analyse the popularity of the account; we considered the parameter number of followers, verified account. Twitter gives an option to "following"; users can follow another user by clicking the \emph{follow} button, and they becomes followers. When a tweet is posted on Twitter, then it is visible to all of his/her followers. Twitter verifies the account, and after doing a verification, Twitter provides the user to receive a blue checkmark badge next to your name. 
From 2017 the service is paused by Twitter, and it is limited to only a few accounts chosen by the Twitter developer. Hence, the verified account is a kind of authentic account. We investigate several characteristics that are associated with popular accounts, namely: Favourites count, follower count, account age and verified status. If a popular user spread false or partially false news, then it is more likely to attract more attention from other users compared to the non-popular twitter handle. 

\subsection{Information diffusion} 
To investigate the diffusion of misinformation i.e, false and partially false tweets, we explore the timeline of retweets and calculate the speed of retweets as a proxy for the speed of propagation. 
A retweet is a re-posting of a tweet, which a Twitter user can do with or without an additional comment. Twitter even provides a retweet feature to share the tweet with your follower network quickly. For our analysis, We only considered the retweet of tweet. 

\paragraph{\textbf{Propagation of misinformation}}
\label{propogationoftweet}

We define the average speed of propagation of tweet as the total number of retweet done for a tweet divided by the total number of days the tweet is getting retweets. The formula to calculate the propagation speed is defined in Equation~\ref{eq1}.

\begin{equation} \label{eq1}
\centering
\begin{split}
P_{s} & = \frac{\sum_{n=1}^{d} rc}{N_{d}} 
\end{split}
\end{equation}

Where P\textsubscript{s} is the propagation speed, rc is retweet count per day and N\textsubscript{d} is the total number of days. 

We calculated the speed of propagation over three different periods. The first metric P\textsubscript{s\_a} is the average overall propagation speed: the speed of retweets from the 1st retweet to the last retweet of a tweet in our data. The second metric is the propagation speed during the peak time of the tweet, denoted by  P\textsubscript{s\_pt}. After a time being, the tweet does not get any retweet, but again some days again start getting user attention and retweet. So, We define the peak time of the tweet as the time (in days) from the retweet start till retweet goes to zero for the first time. The third metric  P\textsubscript{s\_pcv} is the propagation speed calculated during a first peak time of the crisis, i.e., from 15-03-2020 to 15-04-2020. We decided the peak time according to the timeline propagation of retweet, as shown in \ref{fig:timeline}, which is maximum during the mid-March and mid-April. 

Although we are aware that misinformation gets spread on Twitter as soon as it is shared, in the current situation it is not possible us to detect fake tweets in real-time because fact-checking websites take a few days to verify the claim. For example, the fake news on "Has Russia's Putin released lions on streets to keep people indoors amid coronavirus scare?" was first seen on the 22nd March on Twitter but the first fact-checked article published on late 23rd March 2020~\cite{lion} and later by other fact-checking websites. With propagation speed, our aim is to measure the speed of propagation speed among false and partially false tweets.

\subsection{Content analysis}
In order to attain a better understanding of what misinformation around the topic of COVID-19 is circulating on Twitter, we investigate the content of the tweets. Due to the relatively small number of partially false claims, we combined the data for these analyses. First, we analyse the most common hashtags and emojis.
Second, we investigate the most distinctive terms in our data to gain a better understanding of how COVID-19 misinformation differs from other COVID-19 related content on Twitter. To this end, we compare our data to a background corpus of all English COVID-19 tweets from 14-01-2020 to 10-07-2020 (See Section \ref{sec:back}). This enables us to find the most distinctive phrases in our corpus: Which topics are discussed in misinformation that are not discussed in other COVID-19 related tweets? These topics may be of special interest, as there may be little correct information to balance the misinformation circulating on these topics. Third, we make use of the language used in the circulating misinformation to gauge the emotions and underlying psychological factors authors display in their tweets. The latter may be able to give us a first insight into why they are spreading this information. Again the prevalence of emotional and psychological factors is compared to their prevalence in a background corpus in order to uncover how false tweets differ from the general chatter on COVID-19. 

\subsubsection{Hashtags and emojis}
Hashtags are brief keywords or abbreviations prefixed by the hash sign \emph{\#} that are used on social media platforms to make tweets more easily searchable~\cite{Bruns2013}. Hashtags can be considered self-reported topics that the author believes his or her tweet links to. Emoji are standardised pictographs originally designed to convey emotion between participants in text-based conversation~\cite{kelly2015}. Emojis can thus be considered a proxy for self-reported emotions by the author of the tweet. 

We analyse the top 10 hashtags by combining all terms prefixed by a \#. For \# symbols that are stand-alone, we take the next unigram to be the hashtag. We identify emojis using the package emoji~\cite{emojiPython}.

\subsubsection{Analysis of distinctive terms}
To investigate the most distinctive terms in our data, we used the pointwise Kullback Leibner divergence for Informativeness and Phraseness (KLIP)~\cite{tomokiyo-hurst-2003-language} as presented in~\cite{Verberne2016}\footnote{See \url{https://github.com/suzanv/termprofiling} for an implementation.}  for unigrams, bigrams and trigrams. Kullback–Leibler divergence is a measure from information theory that estimates the difference between two probability distributions. The informativeness component (KLI) of KLIP compares the probability distribution of the background corpus to that of the candidate corpus to estimate the expected loss of information for each term. The terms with the largest loss are the most informative. The phraseness component (KLP) compares the probability distribution of a candidate multi-word term to the distributions of the single words it contains. The terms for which the expected loss of information is largest are those that are the strongest phrases. We set the parameter $\gamma$ to 0.8 as recommended for English text. $\gamma$ determines the relative weight of the informativeness component KLI versus the phraseness component KLP. Additionally, for this analysis, tweets that are duplicated after preprocessing are removed; these are slightly different tweets concerning the same misinformation and would bias our analysis of distinctive terms towards a particular claim.

\subsubsection{Analysis of emotional and psychological processes}
The emotional and psychological processes of authors can be studied by investigating their language use. A well-known method to do so is the Linguistic Inquiry and Word Count (LIWC) method~\cite{tausczik2010psychological}. We made use of the LIWC 2015 version and focused on the categories: Emotions, Social Processes, Cognitive Processes, Drives, Time, Personal Concerns and Informal Language. In short, the LIWC counts the relative frequency of words relating to these categories based on manually curated word lists.  All statistical comparisons were made with Mann-Whitney U tests.

\section{Results \label{sec:exp}}

This section describes the result obtained from our analysis for both datasets I, i.e., 1\,500 tweets, which classified as misinformation and dataset II, i.e., 163\,096 COVID-19 tweets as the background tweets. A detailed comparison of two datasets is shown in table ~\ref{tab:datasum}.

\subsection{Account categorisation}
From 1\,500 tweets, we filter 1\,187 unique accounts and performed categorisation of the account using the method mentioned in Section~\ref{account}. The summary of the result obtained is discussed as follow.

\paragraph{\textbf{Bot detection}} From BotoMeter API, we use the Complete Automation Probability (CAP) score to classify the bot. CAP is the probability of the account being a bot according to the model used in the API. We choose the CAP score of more than 0.65. We discovered that there are 24 bot accounts out 1\,187 unique user accounts; user IDs 1025102081265360896 and 1180933034423529473, for instance, are classified as bots.

\paragraph{\textbf{Brand detection}} 
For Brand detection, we used the TwiRole API to categorised the accounts as a brand, male and female. We also randomly checked the category of the account. We have got 792 accounts as a brand. For instance, user ID 18815507 is an organisation account while user ID 2161969770 is a representative of the UNICEF.

\paragraph{\textbf{Popularity of account}} For measuring the popularity, we gathered the information about favourite counts gained by the accounts, followers count, friends accounts, and the age of the accounts using the Twitter API.
We represented the median of Favourite Count, account age and followers count, as shown in Table \ref{tab:datasum}.

\begin{figure*}[pos=ht!]
	\centering
	\includegraphics[width=0.99\textwidth]{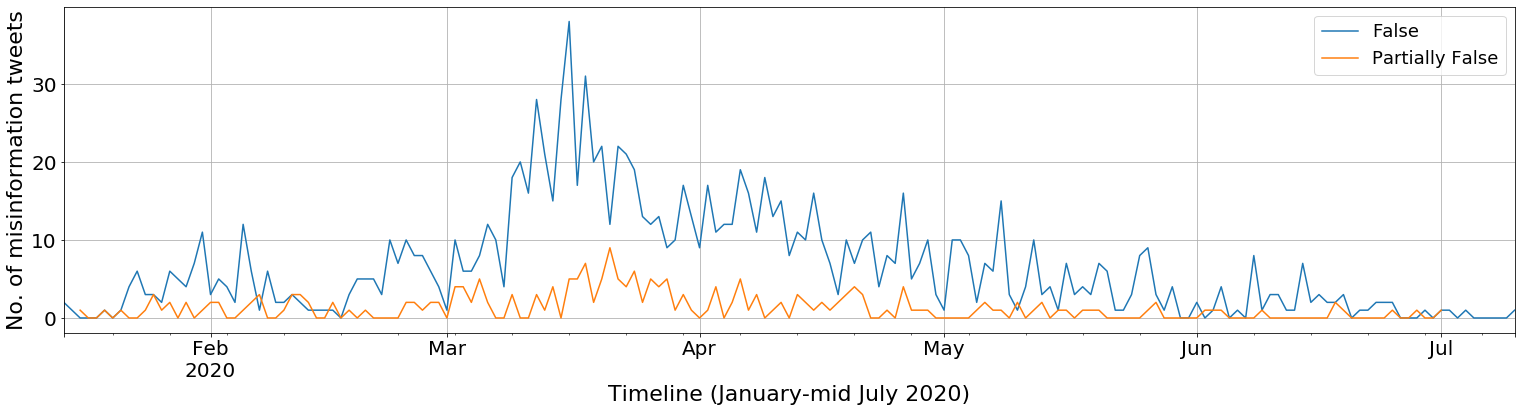}
	\caption{Timeline of misinformation tweets created during January 2020 to mid-July 2020}
	\label{fig:timeline}
\end{figure*}

\begin{table}[pos=ht!]
	\caption{\blue{Description of twitter accounts and tweets from  
			Dataset I(Misinformation) and Dataset II (Background corpus)}}
	\label{tab:datasum}
	\begin{tabular}{@{}p{3.3cm} rr@{}} 
		
		\toprule
		Dataset: & I & II\\   
		
		\bottomrule
		
		Number of Tweets  & 1\,274/276(1\,500)    & 163\,096 \\  
		Unique Account    & 964/198(1\,117) & 143\,905 \\ 
		Verified Account     & 727/131(858) & 16\,720 \\ 
		Distinct Language    & 31/21(33)  &  1(en) \\ 
		Organisation/Celebrity   & 698/135(792)  & 16\,324 \\ 
		Bot  Account  & 22/2(24) &  1\,206\\ 
		Tweet without Hashtags & 919/147(1\,066) & 134\,242\\ 
		Tweet without mentions & 1019/176/(1\,195) & 71\,316\\ 
		Tweet with Emoji & 168/20/(188) & 14\,021\\ 
		
		Median Retweet Count   & 165/169(165)  & 8 \\ 
		Median Favourite  Count   & 2\,446/3\,381(2\,744) & 9\,695 \\ 
		Median Followers Count   & 74\,632/69\,725(74\,131)  & 935 \\ 
		Median Friends Count   & 526/614(531) &  654 \\ 
		Median Account Age (d)   & 82/80(82) & 108                                                       \\ \bottomrule
	\end{tabular}
	
\end{table}

\subsection{Information diffusion}

In this section, we describe the propagation of misinformation with timeline analysis and speed of propagation. 

\paragraph{\textbf{Timeline of misinformation tweets}}
We presented the timeline of the misinformation tweets created during January 2020 to mid-July 2020 in Figure~\ref{fig:timeline}. The blue colour indicates the propagation of the \emph{false} category, whereas orange colour indicates the \emph{partially false} category. The timeline plot of tweet shows that the spread of misinformation of false category is faster than the partially false category. During the peak time of the COVID-19, i.e., mid-March to mid-April, the number of false and partially false tweets were maximum.

\begin{figure*}[pos=tp!]
	\centering
	\includegraphics[width=0.68\linewidth]{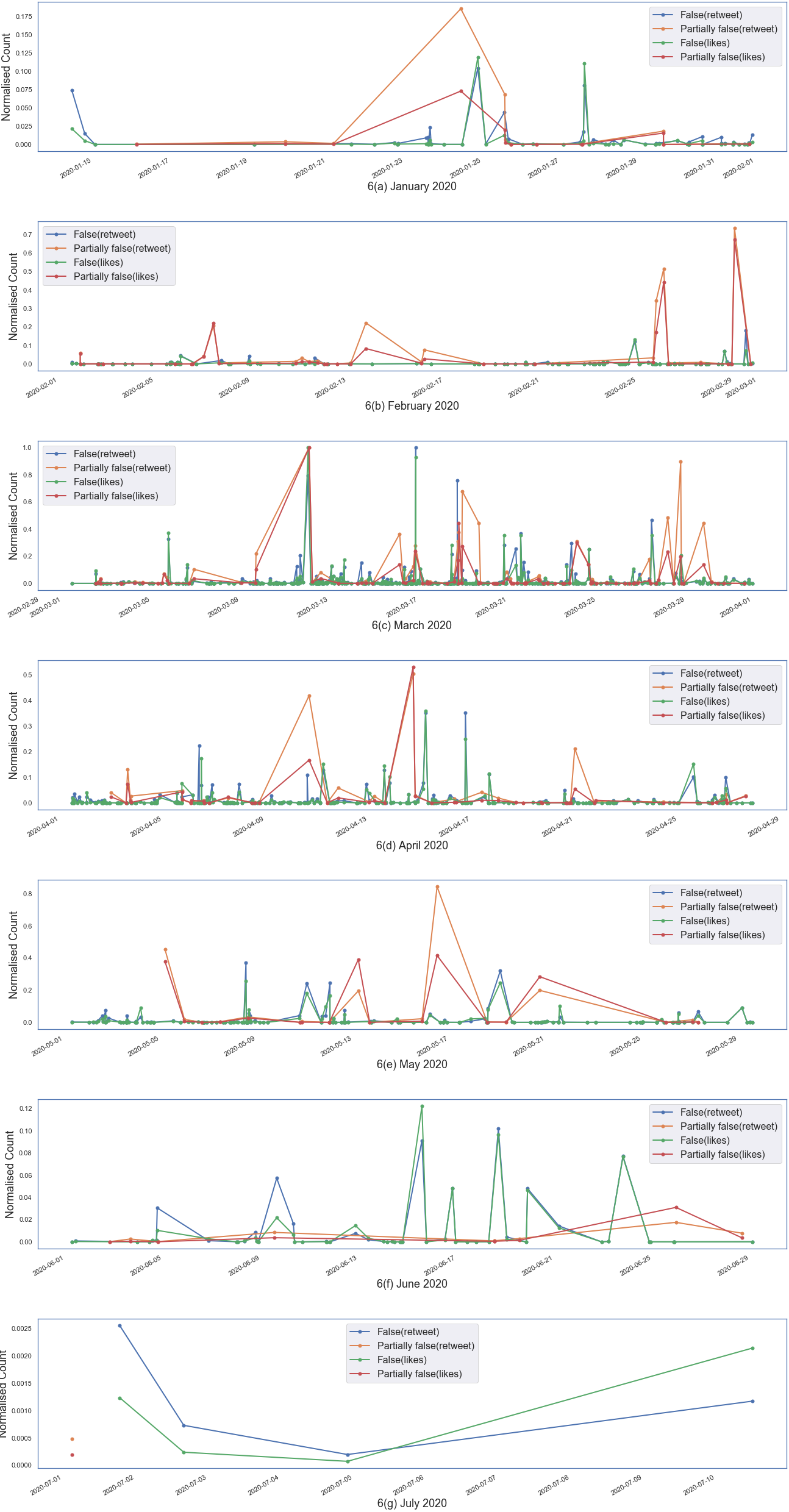}
	\caption{Frequency distribution of retweet(time window of 3 hours) for false (blue) and partially false (orange) claims for each month- 6(a) January, 2020, 6(b) February, 2020, 6(c) March, 2020, 6(d) April, 2020, 6(e) May, 2020, 6(f) June, 2020, 6(g) July, 2020}
	\label{fig:ret}
\end{figure*}

The diffusion of misinformation tweets can be analysed in terms of likes and retweet \cite{stieglitz2013emotions}. Likes indicate how many times a user clicked the tweet as a favourite while retweet is when a user retweets a tweet or retweet with comment. We have visualised the number of likes and retweet gained by each misinformation tweet with the timeline. There is considerable variance in the count of retweet and likes, so we decided to normalise the data. We normalise the count of likes and retweet using Min-Max Normalization in the scale of $[0,1]$. We normalised the count of retweet and likes for the overall month together and plotted the normalised count of retweet and liked for both false and partially false and plotted it for each month. In Figure~\ref{fig:ret} (p.~\pageref{fig:ret}), we presented our result from January to July 2020, one plot for each month. The blue colour indicates the retweet of the \emph{false} category, whereas orange colour indicates the retweet of the \emph{partially false} category. Similarly, the green colour shows the likes of the \emph{false} category, whereas red colour shows the likes of the \emph{partially false} category.

The timeline analysis of normalised retweet shows that misinformation(false category) gets more likes than the partially false category, especially during mid-March to mid-April 2020. The spread of misinformation was at a peak from 16th March to 23rd April 2020, as shown in Figure~\ref{fig:timeline}. However, for the retweet, there is no uniform distinction between false and partially false category. Although, for overall misinformation tweets, the number of likes is comparatively more than the retweet.

\paragraph{\textbf{Propagation of misinformation}}
We calculated the three variant of propagation speed of tweet as discussed in Section~\ref{propogationoftweet}. Results for P\textsubscript{s\_a}, P\textsubscript{s\_pt} and P\textsubscript{s\_pcv} are describe in Table~\ref{tab:propagationSpeed}. We have observed that the speed of propagation is higher for the false category and it was the highest during the peak time of tweet (time duration from the beginning to the day tweet not getting new retweet). 

We performed a chi-square test on the propagation speed shown in table \ref{tab:propagationSpeed}. The analysis showed that there is a difference in the speed of propagation in
tweets, between false and partially false by performing  (X\textsuperscript{2} (3, N = 1500) = 10.23, p <.001). The speed of propagation for the false tweet was more than a partially false tweet, which means the false tweets speed faster. In particular, the propagation speed was maximum during the peak time of the COVID-19 according to Figure~\ref{fig:timeline}.

\begin{table}[width=.99\linewidth,cols=4,pos=ht]
	\caption{Propagation speed of retweet for misinformation tweets}
	\label{tab:propagationSpeed}
	\begin{tabular*}{\tblwidth}{@{} LLLL@{} }
		\toprule
		& False($\sigma$) & Partially False($\sigma$) & Overall($\sigma$)\\
		\midrule
		P\textsubscript{s\_a} & 365(15.6) &  260(6.5) & 209(13.6)  \\  
		P\textsubscript{s\_pt} &  526(26.6)  & 394(14.6) & 376(22.9)  \\ 
		P\textsubscript{s\_pcv} &  418(17.4)  & 357(9.7) & 329(15.2)  \\ 
		
		\bottomrule
	\end{tabular*}
\end{table}

\subsection{Content analysis}

This section discusses the result obtained after doing the content analysis of tweets discussing false and partially false claims.

\subsubsection{Hashtag \& emoji analysis}
\begin{figure}[pos=t]
	\centering
	\includegraphics[width=0.45\textwidth]{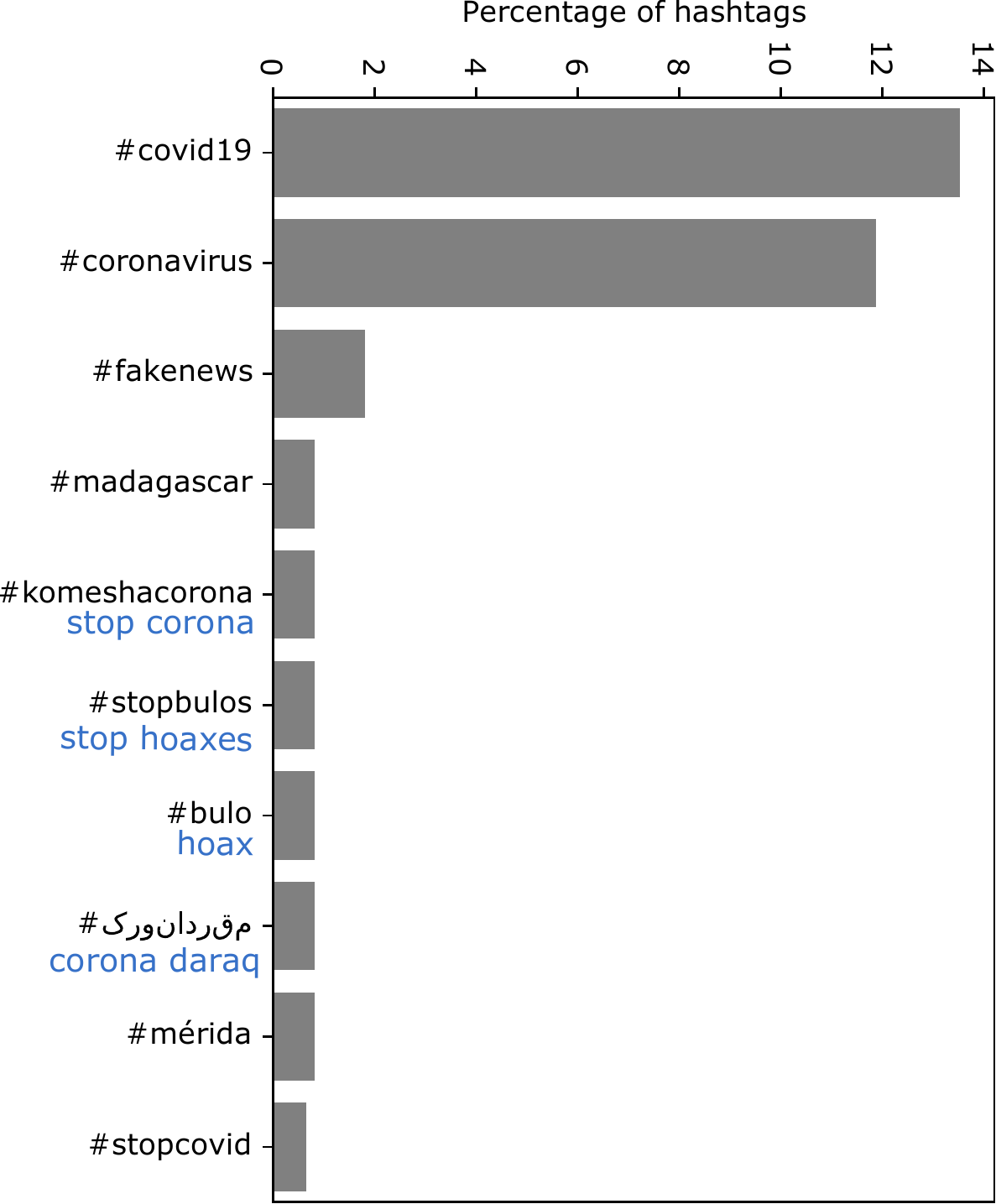}
	\caption{Top 10 hashtags used in the tweets with misinformation. (Translation provided in blue where necessary.)} 
	\label{fig:hash}
\end{figure}

\paragraph{\textbf{Hashtag analysis}} As illustrated in Figure \ref{fig:hash}, many of the most commonly used hashtags in COVID-19 misinformation concern the corona virus itself (i.e. \#covid19 and \#coronavirus) or stopping the virus (i.e. \#stopcorona and \#komeshacorona). Since we did not use any hashtags in the data collection of our corpus of COVID-19 misinformation (See Section \ref{sec:datacol}), this confirms that our method managed to capture misinformation related to the corona crisis. Additionally, the hashtags \#fakenews, \#stopbulos and \#bulo stand out; the author appear to be calling out against other misinformation or hoaxes. The term fake news is widely used to refer to inaccurate information~\cite{Zubiaga2017a} or more specifically to ``fabricated information that mimics news media content in form but not in organisational process or intent''~\cite{Lazer1094}. It appears that some authors are discrediting information spread by others. Yet, we are unable to determine based on this analysis who they are discrediting. Furthermore, three locations can be discerned from the hashtags: Madagascar, M\'{e}rida in Venezuela, and Daraq in Iran. Manual analysis revealed that various false claims are linked to each location. For example, there was misinformation circulating on Facebook that the Madagascan president was calling up African states to leave the WHO \cite{exampleMadagascar1} and the Madagascan government has been promoting the disputed COVID-19 organics, a herbal drink that could supposedly cure an infection~\cite{exampleMadagascar2}. M\'{e}rida is mainly mentioned in relation to a viral image in which one can see a lot of money lying on the street. On social media, people claimed that Italians were throwing money onto the streets to demonstrate that health is not bought with money, while in reality bank robbers had dropped money on the streets in M\'{e}rida, Venezuela \cite{AfricaCheck2020}. Iran has also been central to various false claims, such as the claim by the head of judiciary in Iran that the first centres that volunteered to suspend activities in response to the pandemic were religious institutions \cite{exampleDaqar}, implying that the clergy were pioneers in trying to halt the spread 

\begin{figure}[pos=t]
	\centering
	\includegraphics[width=0.47\textwidth]{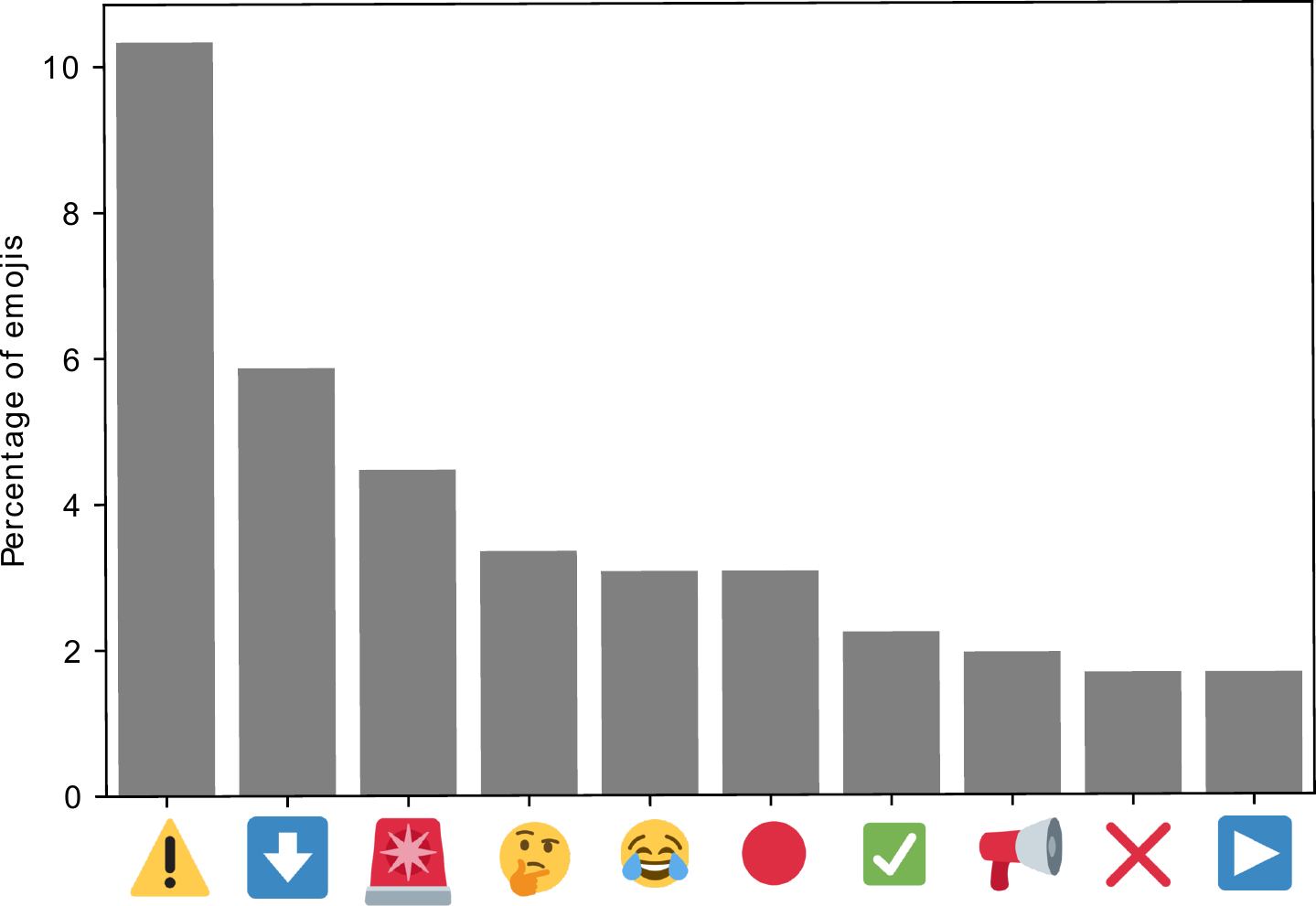}
	\caption{Top 10 emojis used in the tweets}
	\label{fig:emoji}
\end{figure}

\paragraph{\textbf{Emoji analysis}} Emojis are used on Twitter to convey emotions. We analysed the most prevalent emojis used by authors of COVID-19 misinformation on Twitter (see Figure~\ref{fig:emoji}). It appears authors make use of emojis to attract attention to their claim (loudspeaker, red circle) and to convey distrust or dislike (down-wards arrow, cross) or danger (warning sign, police light). In our data set, the thinking emoji is mostly used as an expression of doubt, frequently relating to whether something is fake news or not. Tweets with the laughing emoji are either making fun of someone (e.g. the WHO director-general) or laughing at how dumb others are. Both the play button and the tick are often used for check lists, although the latter is occasionally also conveying approval.

\subsubsection{Most distinctive terms in COVID-19 misinformation}
Analysing the most distinctive terms in our corpus compared to a corpus of general COVID-19 tweets can reveal which topics are most unique. The more unique a topic is to the misinformation corpus, the more likely it is that for this topic there is a larger amount of misinformation than correct information circulating on Twitter. 
We can see in Table \ref{tab:inf} which phrases have the highest KLIP score and thus are most distinct to our corpus of COVID-19 misinformation when compared to the background corpus of COVID-19 tweets. First, we find that misinformation more often concerns discrediting information circulating on social media (`fake news', `circulating on social', `social network', `social media' and `circulating'). Second, compared to general COVID-19 tweets, completely false misinformation more often mentions governing bodies related to health (`world health organisation' and `ministry of health') and their communication to the outside world (`medium briefing'). Conversely, partially false misinformation appears more concerned with human-to-human transmission, mortality rates and running updates on a situation (`latest information' and `situation report') than the average COVID-19 tweet. It is interesting that also the term `mild criticism' is distinctive of partially false claims; It is congruent with the idea that these authors agree with some and disagree with other correct information. Manual inspection reveals that both the terms `bay area' and `santa clara' refer to a serology (i.e. antibody prevalence) study on COVID-19 that was done in Santa Clara, California. Join the homage, and several voitur most likely have a high KLIP score due to the combination of translation errors and the fact that non-English words occur only in the misinformation corpus (also see Section \ref{sec:limit}).

\begin{table}[pos=t!]
	\centering
	\caption{Top 10 most informative terms in misinformation tweets compared to COVID-19 background corpus}
	\label{tab:inf}
	\begin{tabular}{@{}ll@{}}
		\toprule
		False Claims             & Partially False Claims      \\ \midrule
		social medium             & fake news                   \\
		circulating on social     & mortality rate              \\
		social network            & mild criticism              \\
		fake news                 & join the homage             \\
		not                       & several voitur              \\
		corona virus              & human-to-human transmission \\
		medium briefing           & bay area                    \\
		world health organization & santa clara                 \\
		ministry of health        & latest information          \\
		circulating               & situation report            \\ \bottomrule
	\end{tabular}
	\vspace{-1em}
\end{table}

\subsubsection{Psycho-linguistic analysis}

According to Wardle and Derakhshan \cite{Wardle2017}, the most successful misinformation plays into people's emotions. To investigate the emotions and psychological processes portrayed by authors in their tweets, we use the LIWC to estimate the relative frequency of words relating to each category~\cite{tausczik2010psychological}. LIWC is a good proxy for measuring emotions in tweets: In a recent study of emotional responses to COVID-19 on Twitter, Kleinberg \textit{et al.}~\cite{Kleinberg2020} found that the measures of the LIWC correlate well with self-reported emotional responses to COVID-19.

\begin{figure*}[pos=ht!]
	\centering
	\includegraphics[width=0.59\textwidth]{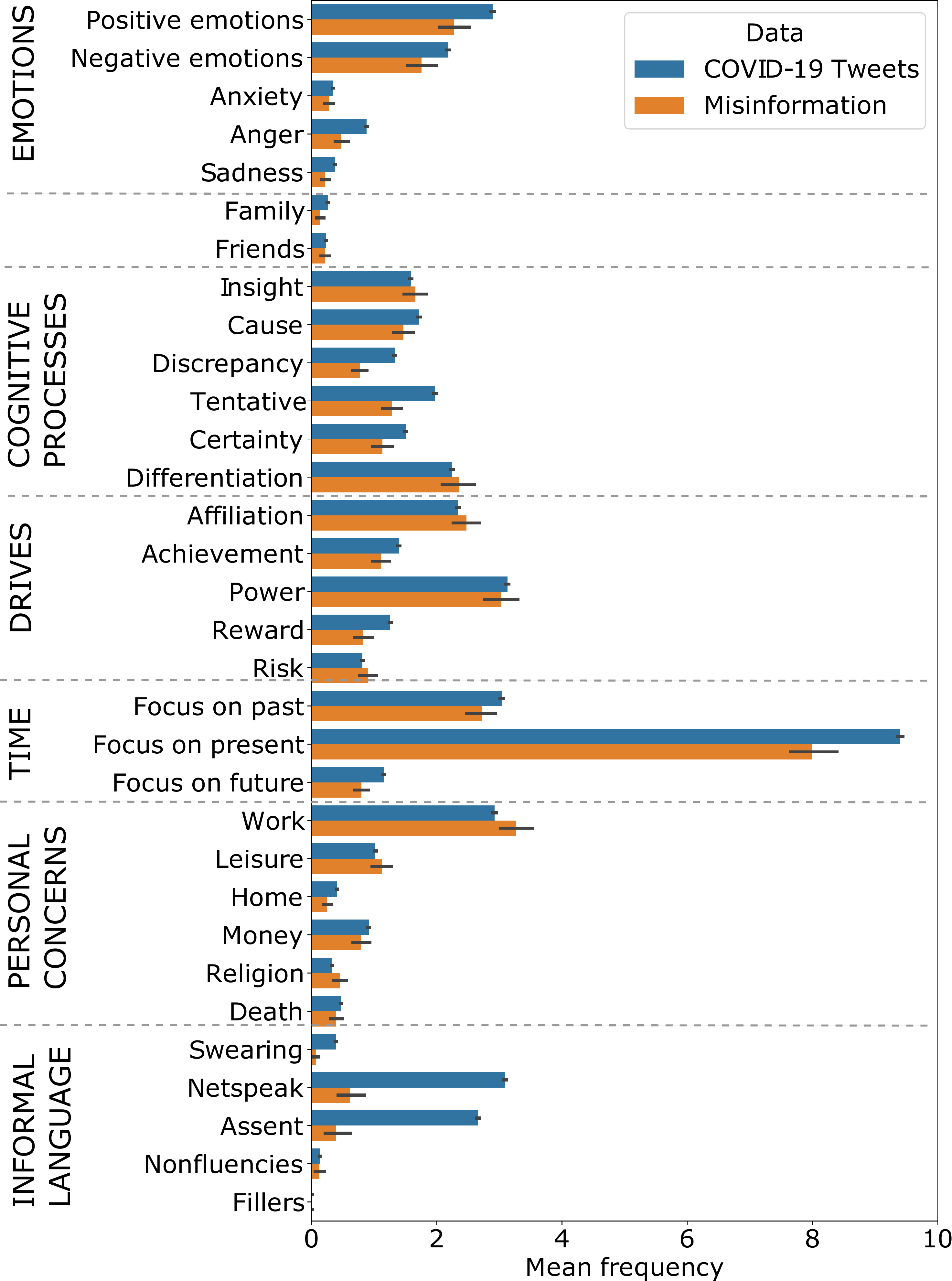}
	\caption{LIWC results comparing COVID-19 background corpus to COVID-19 misinformation}
	\label{fig:LIWCResultsBackground}
	\vspace{-0.4em} 
\end{figure*}

First, we compared the false with the partially false misinformation, but there do not seem to be significant differences in language use. Second, we compared all misinformation on COVID-19 with a background corpus of tweets on COVID-19. Both positive and negative emotions are significantly less prevalent in tweets with COVID-19 misinformation than in COVID-19 related tweets in general ($p<0.0001$) (Figure~\ref{fig:LIWCResultsBackground}). This is also the case for specific negative emotions such as anger and sadness ($p<0.0001$ and $p= 0.0002$ resp.). The levels of anxiety expressed are not significantly different, however. 

When we consider cognitive processes that can be discerned from language use, we see that authors of tweets containing misinformation are significantly less tentative in what they say ($p<0.0001$) although they also contain less words that reflect certainty ($p<0.0001$). Moreover, they contain less words relating to the discrepancy between the present (i.e. what is now) and what could be (i.e. what would, should or could be) ($p<0.0001$). 

We also consider indicators of what drives authors. It appears that authors posting misinformation are driven by affiliations to others ($p=0.003$) significantly more often than authors of COVID-19 tweets in general and significantly less often by rewards ($p<0.0001$) or achievements ($p= 0.001$). In line with the lack of focus on rewards, money concerns also appear to be discussed less ($p<0.0001$). Thus, authors posting misinformation appear to be driven by their desire to prevent people they care about from coming to harm. Interestingly, tweets on misinformation are significantly less likely to discuss family ($p<0.0001$) although not less likely to discuss friends. 

Misinformation is also less likely to discuss certain personal concerns, such as matters concerning the home ($p=0.002$) or money ($p<0.0001$) but also death ($p=0.03$). Yet, they are more likely to discuss personal concerns relating to work ($p<0.0001$) and religion ($p<0.0001$).

Furthermore, COVID-19 tweets appear to have a focus on the present. COVID-19 misinformation seems to also focus on the present but to a significantly lesser degree ($p<0.0001$). Tweets containing misinformation are also less focused on the future ($p<0.0001$), whereas the focus on past does not differ. Lastly, although both corpora are from Twitter, the COVID-19 tweets containing misinformation use relatively less informal language with less so-called netspeak (e.g. lol and thx) ($p< 0.0001$), swearing ($p<0.0001$) and assent (e.g. OK) ($p< 0.0001$). The smaller amount of assent words might also indicate that these tweets are expressing a disagreement with circulating information, in line with the results from the emoji analysis.

\section{Discussion \label{sec:dis}}
Based on our analysis, we discuss our findings.
We first look at lessons learned from using Twitter in an ongoing crisis before deriving recommendations for practice. We then scrutinise the limitations of our work, which form the basis for our summary of open questions.

\subsection{Lessons learned}
While conducting this research, we encountered several issues concerning the use of Twitter data to monitor misinformation in an ongoing crisis. We wanted to point these out in order to stimulate a discussion on these topics within the scientific community. 

The first issue is that the Twitter API severely limits the extent to which the reaction to and propagation of misinformation can be researched after the fact. One of the major challenges with collecting Twitter data is the fact that the Twitter API does not allow for retrieval of tweet replies over 7 days old and limits the retrieval of retweets. As it typically takes far longer for a fact-checking organisation to verify or discount a claim, this means early replies cannot be retrieved in order to gauge the public reaction before fact-checking. Recently, Twitter has created an endpoint specifically for retrieving COVID-19 related tweets in real-time for researchers \cite{Twitterdev}. Although we welcome this development, this does not solve the issue at hand. Although large data sets of COVID-19 Twitter data are increasingly being made publicly available \cite{chen2020covid19}, as far as we are aware, these do not include replies or retweets either. 

The second issue is that there is an inherent tension between the speed at which data analysis can be done to aid practitioners combating misinformation and the magnitude of Twitter data that can be included. In a crisis where speed is of the essence, this is not trivial. Our data was limited by the number of claims that included a tweet (for more on data limitations see Section \ref{sec:limit}), causing a loss of around 90\% of the claims we collected from fact-checking websites. This problem could be mitigated to some extent by employing similarity matching to map misinformation verified by fact-checking organisations to tweets in COVID-19 Twitter data \cite{chen2020covid19}. However, this would be computationally intensive and require the creation of a reliable matching algorithm, making this approach far slower. Moreover, automatic methods for creating larger data sets will also lead to more noisy data. Thus, such an approach should rather be seen as complementary to our own.
Probably, social media analytics support can draw from lessons learned on crisis management decision making under deep uncertainty~\cite{IJEM2019}.
Eventually, more work and scientific debate on this topic are necessary. Additionally, as an academic community, it is important to explicitly convey what can and what cannot be learned from the data so as to prevent practitioners from drawing unfounded conclusions.
The other way around, we deem it necessary to ``look over the shoulder'' over practitioners to learn about their way of handling the dynamics of social media, eventually leading to a better theory.

A third point that must be considered by the academic community researching this subject is the risk of profiling Twitter users. There have been indications that certain user characteristics such as gender \cite{chen2015students} and affiliation with the alt-right community \cite{TheGuardian} may be related to the likelihood of spreading misinformation. Systematic analyses of these characteristics could prove valuable to practitioners battling this infodemic but simultaneously raises serious concerns related to discrimination. In this article, we did not analyse such characteristics, but we urge the scientific community to consider how this could be done in an ethical manner. Better understanding which kind of people create, share and succumb to misinformation would much help to mitigate their negative influence.

Fourth, relying on automatic detection of fact-checked articles can lead to false results. The method used by fact-checkers is often confusing and messy— it is a muddle of claims, news articles and social media posts. Additionally, each fact-checker appears to have its own process of debunking and set of verdicts. We even encountered cases where fact-checkers discuss multiple claims in one go, resulting in additional confusion. Moreover, fact-checkers do not always explicitly specify the final verdict or class (false or not) of the claim. For example, in a fact check performed by Pesa Check~\cite{pesacheck} the claim "Chinese woman was killed in Mombasa over COVID-19 fears" is described and the article links various news sources. Then, abruptly in the bottom, a tweet message about a mob lynching of a man is embedded, and no specification of the class (false or not) of the article is mentioned. 

\subsection{Recommendations}
Research on a topic that relates to crisis management offers the chance to not only contribute to the scientific body of knowledge but directly (back) to the field.
Our findings allow us to draw the first set of recommendations for public authorities and others with an \emph{official} role in crisis communication.
Ultimately, these also could be helpful for all critical users of social media, and especially those who seek to debunk misinformation.
\blue{While we are well aware that we are unable to provide sophisticated advice well-backed by theory, we believe our recommendations may be valuable for practitioners and can lead the way to contributions to theory. In fact, public authorities seek for such advice based on scientific work, which can be the foundation for local strategies and aid in arguing for measurements taken
	(cf. e.g. with the work presented by \citet{ISCRAM2017Portal,HICSS2018SMR}).}

First, and rather unsurprisingly, closely watching social media is recommended (cf. e.g. with \cite{alexander2014social,wukich2015social,veil2011work}).
COVID-19 has sparked much misinformation, and it quickly propagates.
Our work indicates that this is not an ephemeral phenomenon.
For the \emph{john doe user}, our findings suggest to always be critical, even if alleged sources are given, and even if tweets are rather old or make the reference of old tweets.

Second, our results serve as proof that brands (organisations or celebrities) are involved in approximately 70\% of false category and partially false category of misinformation. They either create or circulate misinformation by performing activities such as liking or retweeting. This is in line with work by researchers from the Queensland University of Technology who also found that celebrities are so-called ``super-spreaders'' of misinformation in the current crisis~\cite{Brisbane}. Thus, we recommend close monitoring of celebrities and organisations that have been found to spread misinformation in order to catch misinformation at an early stage.
For users, this means that they should be cautious, even if a tweet comes from their favourite celebrity.

Third, we recommend close monitoring of tags such as \#fakenews that are routinely associated with misinformation. For Twitter users, this also means that they have chances to check if a tweet might be misinformation by checking replies to it -- these replies being tagged with for instance \#fakenews would be an indicator of suspicion \footnote{This also applies the other way around: facts might be commented ironically or deliberately provocative by using the tag \#fakenews to create confusion and to make trustworthy sources appear biased.}.

Fourth, we advise to particularly study news that are partially false -- despite an observed slower propagation, it might be more dangerous to those not routinely resorting to information that provides the desired reality rather than facts. As mentioned before, it may be more challenging for users to recognise claims as false when they contain elements of truth, as this has found to be the case even for professional fact-checkers~\cite{Lim2018}. It is still an open question whether there is less partially false than false information circulating on Twitter or whether fact-checkers are more likely to debunk completely false claims.

Fifthly, we recommend authorities to carefully tailor their online responses.
We found that for spreading fake news, emojis are used to appeal to the emotions.
One the one hand, you would rather expect a trusted source to have a neutral, non-colloquial tone. On the other hand, it seems to advisable to get to the typical tone in social media, e.g. by also using emojis to some degree. 
We also advise authorities to employ tools of social media analytics.
This will help them to keep updated on developing misinformation, as we found that for example, psycho-linguistic analysis can reveal particularities that differ in misinformation compared to the ``usual talk'' on social media.
Debunking fake news and keeping information sovereignty is an arms race -- using social media analytics to keep pace is therefore advisable.
In fact, we would recommend authorities to employ methods such as the ones discussed in this paper as they work not only ex-post but also during an ongoing infodemic.
However, owing to the limitation of data analysis and regarding API usage (cf. with the prior and with the following section), we recommend making social media monitoring part of the communication strategy, potentially also manually monitoring it.
This advice, in general, applies to all Twitter users: commenting something is like shouting out loudly on a crowded street, just that the street is potentially crowded by everyone on the planet with Internet access. Whatever is tweeted might have consequences that are unsought for (cf. e.g. \cite{10.1111/jcom.12259}).

Lastly, we recommend working timely, yet calmly, and with an eye for the latest developments.
During our analysis, we encountered much bias -- not only on Twitter but also on media and even in science.
Topics such as the justification of lockdown measurement spark heated scientific debate already and offer much controversy.
Traditional media, which supposedly should have well-trained science journalists, will cite vague and cautiously phrases ideas from scientific preprints as seeming facts, ignoring that that ongoing crisis mandates them to be accessible before peer review.
Acting cautiously on misinformation will not only likely create more misinformation, but it may erode trust.
Our final recommendation for officials is, thus, to be the trusty source in an ocean of potential misinformation.

These recommendations must not be mistaken for definite guidelines, let alone a handbook.
They should offer some initial aid, though.
Moreover, formulating them supports the identification of research gaps, as will be discussed along with the limitations in the following two subsections.

\subsection{Limitations \label{sec:limit}}
Due to its character as complete yet early research, our work is bound to several limitations. Firstly, we are aware that there may be a selection bias in the collection of our data set as we only consider rumours that were eventually investigated by a fact-checking organisation. Thus, our data probably excluded less viral rumours. Additionally, we limited our analysis to Twitter, based on prior research by~\cite{Cinelli} that found that out of the mainstream media, it was most susceptible to misinformation. Nonetheless, this does limit our coverage of online COVID-19 misinformation. 

We are also aware that we introduce another selection bias through our data collection method, as we only include rumours for which the fact-checking organisation refers to a specific tweet id in its analysis of the claim. Furthermore, we cannot be certain that this tweet id refers to the tweet spreading misinformation, as it could also refer to a later tweet refuting this information or an earlier tweet spreading correct information that was later re-purposed for spreading misinformation. Two examples of this are: (1) ``Carnival in Bahia - WHO WILL ? -URL-'' which refers to a video showing Carnival in Bahia but of which some claim it shows a gay party in Italy shortly before the COVID-19 outbreak~\cite{AFP2020} and (2) 
`` i leave a video of what happened yesterday 11/03 on a bicentennial bank in merida. Yes, these are notes.'' which is the correct information for a video from 2011 that was re-purposed to wrongly claim that Italians were throwing cash away during the corona crisis~\cite{AfricaCheck2020}. 

Second, our interpretation of both hashtag and emoji usage by authors of misinformation is limited by our lack of knowledge of how the authors intended them. Both are culturally and contextually bound, as well as influenced by age and gender~\cite{Herring2020} and open to changes in their interpretation over time~\cite{Miller2016}. 

Thirdly, despite indications of a rapid spread of COVID-19 related misinformation on Twitter, a recent study by Allen \textit{et al.} \cite{Allen2020} found that exposure to fake news on social media is rare compared to exposure to other types of media and news content, such as television news. They do concede that it may have a disproportionate impact, as the impact was not measured. Regardless, further research not only into the prevalence but also into the exposure of people to COVID-19 misinformation is necessary. Our study does not allow for any conclusions to be drawn on this matter.

A fourth limitation stems from the selection of our background corpus. Although the corpora span the same time period, COVID-19 misinformation was allowed to be written in another language than English, whereas we limited our background corpus to English tweets. Therefore, the range of topics discussed in both corpora may also differ for this reason. Consequently, we need to remain critical of informative terms of the COVID-19 misinformation corpus that refer to non-English speaking countries or contain non-English words (e.g. voitur).

However, none of these limitations impairs the originality and novelty of our work; in fact, we gave first recommendations for practitioners and are now able to propose directions for future research.

\subsection{Open questions}
On the one hand, work on misinformation in social media is no new emergence.
On the other hand, the current crisis has made it clear how harmful misinformation is.
Obviously, strategies to mitigate the spread of misinformation are needed.
This leads to open research questions, particularly in light of the limitations of our work.
Open questions can be divided into four categories.

First, techniques, tools and theory from social media analytics must be enhanced.
It should become possible -- ideally in half- or fully-automated fashion -- to assess the propagation of misinformation.
Understanding where misinformation originates, in which networks in circulates, how it is spread, when it is debunked, and what the effects of debunking are ought to be researched in detail.
As we already set out in this paper, it would be ideal for providing as much discriminatory power as possible, for example by distinguishing misinformation that is completely and partly false; misinformation that is spread intentionally and by accident (possibly even with good intention, but not knowing better); and misinformation that is shared only in silos versus misinformation that leaves such silos and propagates further. Not only such a typology (maybe even taxonomy) would make a valuable contribution to theory but also in-depth studies of the propagation by type.
Such insights would also aid fact-checkers, who would, for example, learn when it makes sense to debunk facts, and whether there is a ``break-even'' point after which it is justified to invest the effort for debunking.

Second, since a holistic approach is necessary to effectively tackle misinformation, it is important to investigate how our results -- and future results on the propagation of misinformation on Twitter -- relate to other social media.
While Twitter is attractive for study and important for misinformation due to the brevity and speed, other social media should also be researched. 
COVID-19 misinformation is not necessarily restricted to a single platform and may thus be spread from one platform to another.
Consequently, the fact-checking organisation may not mention any tweets despite a claim also being present on Twitter.
Especially if the origin of the claim was another platform, there might be several seeds on Twitter as people forward links from other platforms.
As part of this, the spread of fake news through closed groups and messages would make an interesting object of study.

Third, the societal consequences of fake news ought to be investigated.
There is no doubt society is negatively impacted, but to which extent these occur, whom they affect, and how the offline spread of misinformation can be mitigated remain open research questions.
Again, achieving high discriminatory power would be much helpful to counter misinformation.
For example, it would be worthwhile to investigate how the diffusion of misinformation about COVID-19 differs per country. 
In this regard, specifically, the relation between trust and misinformation is a topic that requires closer investigation. In order for authorities to maintain information sovereignty, users -- in this case typically citizens -- need to trust the authorities. Such trust may vary widely from country to country. In general, a high level of trust, as achieved in the Nordic countries \cite{listhaug2008trust, andreasson2017trust}, should help mitigating misinformation. Thus, a better understanding of how authorities can gain and maintain a high level of trust could greatly benefit effective crisis management.

Fourth, researching synergetically between the fields of social media analytics and crisis management could benefit both fields. On the one hand, social media analytics could benefit from the expertise of crisis managers and researchers in the field of crisis management in order to better interpret their findings and to guide their research into worthwhile directions.
On the other hand, researchers in crisis management could make use of novel findings on the propagation of misinformation during the crisis to improve their existing theoretical models to provide holistic approaches to information dissemination throughout the crisis. Crisis management in practice needs a set of guidelines. What we provided here is just a starting point; an extension requires additional quantitative and especially qualitative research as well as validation by practitioners. Further collaboration of these fields is necessary.  

\section{Conclusion \label{sec:conclusion}}
In this article, we have presented work on COVID-19 misinformation on Twitter. We have analysed tweets that have been fact-checked by using techniques common to social media analytics. However, we decided on an exploratory approach to cater to the unfolding crisis. While this brings severe limitations with it, it also allowed us to gain insights otherwise hardly possible. Therefore, we have presented rich results, discussed our lessons learned, have first recommendations for practitioners, and raised many open questions. That there are so many questions -- and thereby research gaps -- is not surprising, as the COVID-19 crisis is among few \emph{stress}-like disasters where misinformation is studied in detail\footnote{One other is the European refugee crisis \cite{roozenbeek2019fake}. However, due to the limited time of onset, with regard to crisis management particularly \emph{shocks} -- such as earthquakes and floods -- are studied.} -- and we are just at the beginning. Therefore, it was our aspiration to contribute to a small degree to mitigating this crisis.

We hope that our work can stimulate the discussion and lead to discoveries from other researchers that make social media a more reliable data source.
Some of the questions raised will also be on our future agendas.
We intend to continue the very work of this paper, even though in a less exploratory fashion.
Rather, we will seek to verify our early findings quantitatively with much larger data sets. We will seek collaboration with other partners to gain access to historical Twitter data in order to investigate all replies and retweets to the tweets on our corpus. 
This extension should cover not only additional misinformation but also full sets of replies and retweets.
Moreover, it would be valuable to longitudinally study how misinformation propagates as the crisis develops.
Regarding COVID-19, medical researchers warn of the second wave \cite{xu2020beware}, and maybe consecutive further ones.
Will misinformation also come in waves, possibly in conjunction with societal discussion, political measurements, or other influencing factors?
Besides an extension of the data set, our work will be extended methodologically.
For example, we seek to stance detection methods to determine the position of replies towards the claim. 
At the same time, we would like to qualitatively explore the rationale behind our observation.

Right as we concluded our work on this article, ``doctors, nurses and health expert [\ldots] sound the alarm'' over a ``global infodemic, with viral misinformation on social media threatening lives around the world'' \cite{AlarmInfodemic}.
They target tech companies, specifically those that run social media platforms.
We take their letter as encouragement.
The companies might be able to filter much more misinformation than they do now, but to battle, this infodemic much more is needed.
We hope we could help to arm those that seek for truth!

\printcredits

\bibliography{MisinformationCovid}
\bibliographystyle{cas-model2-names}

\newpage 
\section*{Author biographies}
\bio{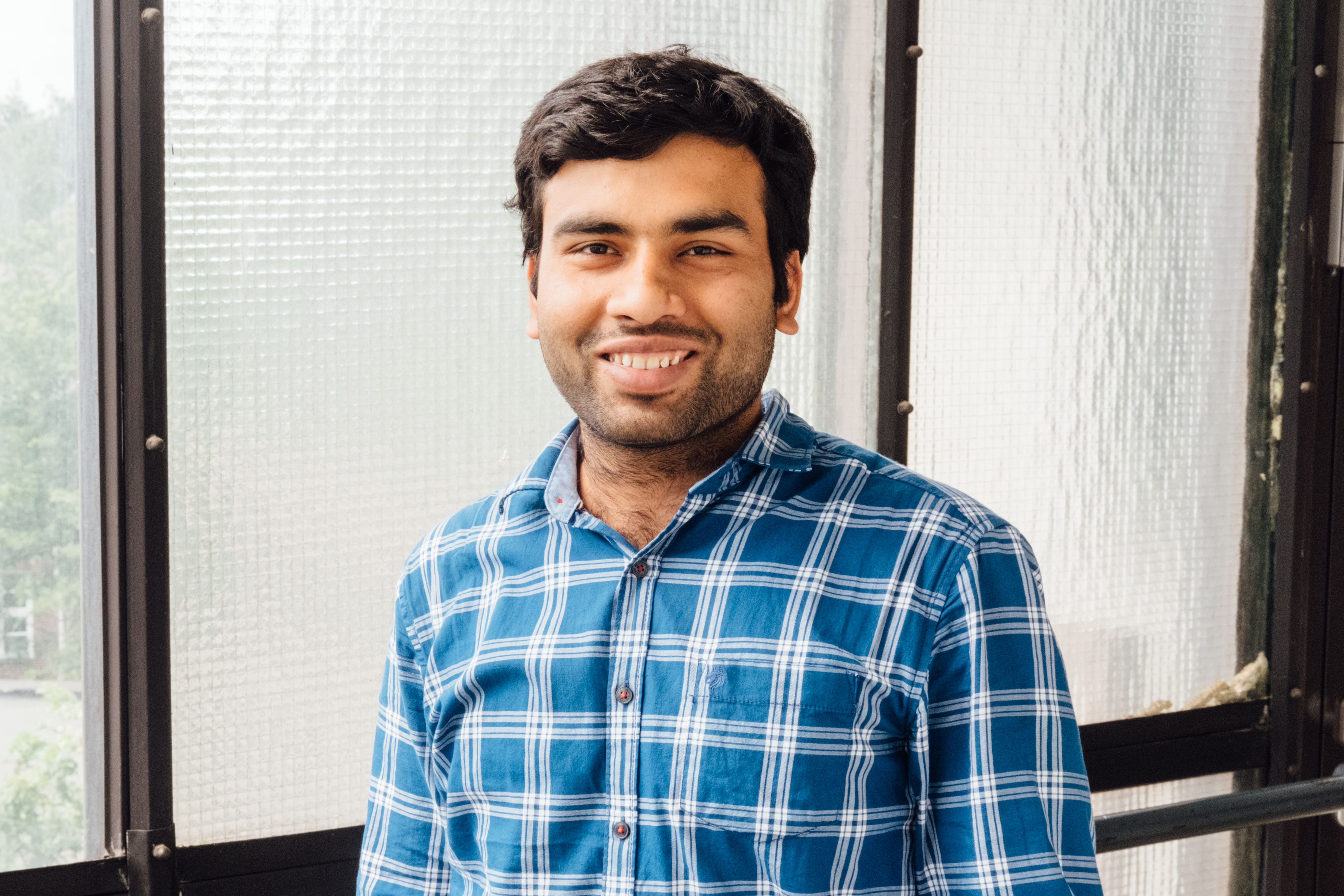}
Gautam Kishore Shahi is a PhD student in the Research Training Group User-Centred Social Media at the research group for Professional Communication in Electronic Media/Social Media(PROCO) at the University of Duisburg-Essen, Germany. His research interests are Web Science, Data Science, and Social Media Analytics. He has a background in computer science, where he has gained valuable insights in India, New Zealand, Italy and now Germany. Gautam received a Master's degree from the University of Trento, Italy and Bachelor Degree from BIT Sindri, India. Outside of academia, He worked as an Assistant System Engineer for Tata Consultancy Services in India.
\endbio

\bio{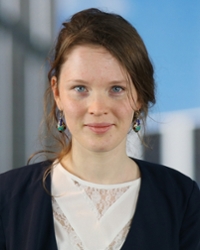}
Anne Dirkson is a PhD student at the Leiden Institute of Advanced Computer Science (LIACS) of the University of Leiden, the Netherlands. Her PhD focuses on knowledge discovery from health-related social media and aims to empower patients by automatically extracting information about their quality of life and knowledge that they have gained from experience from their online conversations. Her research interests include natural language processing, text mining and social media analytics. Anne received a BA in Liberal Arts and Sciences at the University College Maastricht of Maastricht University and a MSc degree in Neuroscience at the Vrije Universiteit, Amsterdam.  
\endbio

\bio{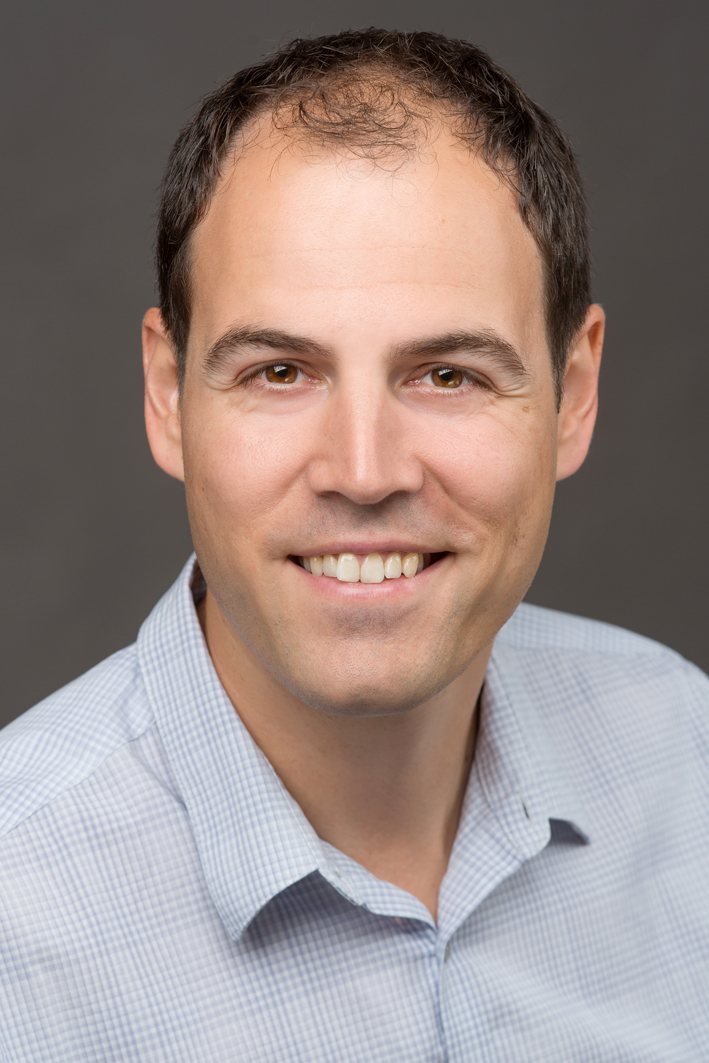}
Tim A. Majchrzak is professor in Information Systems at the University of Agder (UiA) in Kristiansand, Norway. He also is a member of the Centre for Integrated Emergency Management (CIEM) at UiA. Tim received BSc and MSc degrees in Information Systems and a PhD in economics (Dr. rer. pol.) from the University of Münster, Germany. His research comprises both technical and organizational aspects of Software Engineering, typically in the context of Mobile Computing. He has also published work on diverse interdisciplinary Information Systems topics, most notably targeting Crisis Prevention and Management. Tim's research projects typically have an interface to industry and society. He is a senior member of the IEEE and the IEEE Computer Society, and a member of the Gesellschaft für Informatik e.V.
\endbio

\end{document}